\documentclass[pre,aps,twocolumn,superscriptaddress,floatfix]{revtex4}

\usepackage{pdfpages}
\usepackage{amsmath}
\usepackage{amssymb}
\usepackage{amsbsy}
\usepackage{graphicx}
\usepackage{color}
 
\usepackage{enumerate}
\usepackage{mathtools}

\def\be{\begin{equation}}
\def\ee{\end{equation}}
\def\bfi{\begin{figure}}      
\def\efi{\end{figure}}
\def\bea{\begin{eqnarray}}
\def\eea{\end{eqnarray}}

\begin{document}

\title{Controlled viscosity in dense granular materials}

\author{A. Gnoli}
\affiliation{Institute for Complex Systems - CNR, P.le Aldo Moro 2, 00185, Rome, Italy}
\affiliation{Department of Physics, University of Rome Sapienza, P.le Aldo Moro 2, 00185, Rome, Italy}

\author{L. de Arcangelis}
\affiliation{Department of Industrial and Information Engineering, University of Campania ``Luigi Vanvitelli'',  Aversa (CE), Italy}

\author{F. Giacco}
\affiliation{Department of Mathematics and Physics, University of Campania ``Luigi Vanvitelli'', Caserta, Italy}

\author{E. Lippiello}
\affiliation{Department of Mathematics and Physics, University of Campania ``Luigi Vanvitelli'', Caserta, Italy}

\author{M. Pica Ciamarra}
\affiliation{CNR-SPIN, Department of Physics, University ``Federico II'', Naples, Via Cintia, 80126 Napoli, Italy}
\affiliation{Division of Physics and Applied Physics, School of Physics and Mathematical Sciences,
Nanyang, Technological University, 21 Nanyang Link, Singapore, 637371}

\author{A. Puglisi}
\affiliation{Institute for Complex Systems - CNR, P.le Aldo Moro 2, 00185, Rome, Italy}
\affiliation{Department of Physics, University of Rome Sapienza, P.le Aldo Moro 2, 00185, Rome, Italy}

\author{A. Sarracino}
\affiliation{Institute for Complex Systems - CNR, P.le Aldo Moro 2, 00185, Rome, Italy}
\affiliation{Department of Physics, University of Rome Sapienza, P.le Aldo Moro 2, 00185, Rome, Italy}

\begin{abstract}
We experimentally investigate the fluidization of a granular material
subject to mechanical vibrations by monitoring the angular velocity of
a vane suspended in the medium and driven by an external motor. On
increasing the frequency we observe a re-entrant transition, as a
jammed system first enters a fluidized state, where the vane rotates
with high constant velocity, and then returns to a frictional state,
where the vane velocity is much lower.  While the fluidization
frequency is material independent, the viscosity recovery frequency
shows a clear dependence on the material, that we rationalize by
relating this frequency to the balance between dissipative and
inertial forces in the system.  Molecular dynamics simulations well
reproduce the experimental data, confirming the suggested theoretical
picture.
\end{abstract}


\maketitle

\emph{Introduction.}-- Granular systems can be found in solid-like
states able to resist applied stresses~\cite{jeager}, and in flowing
fluid-like states~\cite{brill,puglisi}.  The transition between these
regimes is driven by changes in density and in applied
stresses, as well as by changes in applied forcing.  In this
respect, the role of mechanical
vibrations~\cite{falcon,urb2004,huan,johnsonjia2005,DB05,PCCDM07,BJJ08,johnsonsava2008,marchal,griffa,sellerio,dijk,giacco,vanderelst2012,johnson2012,krim,ciamarra_role_2012,griffa2013,GRL:GRL50813,3ddegriffa2014,maksephyschemestry,giaccoprl2015,
  corwin2015,lasta,PADCC15,BITCCA15,gnoli2,WDH16,SZM16} in driving the
unjamming transition is particularly relevant as related to many
phenomena, from avalanche dynamics~\cite{herrmann} and earthquake
triggering~\cite{dearc} in geophysics, to the manufacturing process in
material, food and pharmaceutical industries~\cite{coussot}.  The
influence of applied vibrations on the transition from a solid to
a fluid-like state, and possibly from the fluid to the solid state,
investigated in some numerical simulations~\cite{capozza,giacco}, is
an issue of great practical relevance. Indeed, its understanding might
open the possibility of controlling the frictional resistance of
granular media~\cite{persson,krim,vanossi}.  This problem has been
addressed by Capozza et al.~\cite{capozza,capozza1} on a prototypical
model of particles confined between two rigid substrates in relative
motion~\cite{persson,braun}, the bottom substrates vertically
vibrating.  Their numerical simulations suggest that viscosity is
reduced when the bottom plate vibrates in a range of frequencies, as
rationalized through a general argument based on the reduction of
effective interface contacts in the system.  However, the validity of
this argument lacks experimental verification in real granular media.

\begin{figure}[!b]
\includegraphics[width=0.5\columnwidth,clip=true]{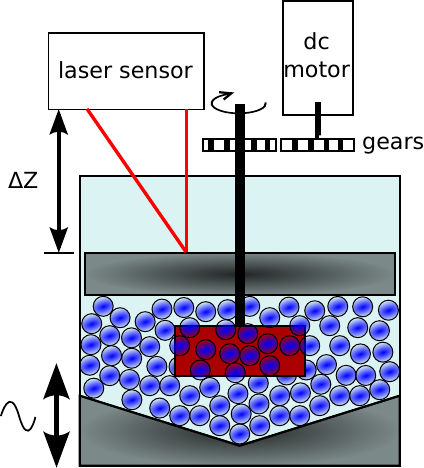}
\caption{Experimental setup. A vane (red rectangle) is
  coupled to a dc motor and is suspended in a dense granular system
  of spherical particles. The container is vertically vibrated
  with sinusoidal oscillations of frequency $f$ and amplitude $A$. On
  the top of the granular medium there is a plate, whose vertical
  displacement $\Delta Z$ is measured with a laser sensor.}
\label{fig00}
\end{figure}

In this Letter we experimentally investigate the fluidization
properties of different granular materials subject to periodic
vertical vibrations, in a wide range of frequencies and amplitudes. We
probe the viscosity features of the granular system by a vane
suspended in it and driven by a motor, see Fig.~\ref{fig00}. Measuring
the average angular velocity of the vane as a function of the
vibration frequency, we are able to explore a broad range of behaviors
of the granular system, from fully jammed to unjammed-fluidized
states.  Our results confirm the existence of a frequency range in
which the system is fluidized, the vane rotating with a finite speed.
The transition from the solid to the fluid state occurs at a frequency
which is very well estimated by the theory of
Ref.~\cite{capozza}, that we confirm also investigating particles of
different materials.  Conversely, we show that the viscosity
  recovery frequency, where the system transitions from the fluid to
the highly viscous state, depends on the material properties. We argue
that the material dependence of the viscosity recovery transition
originates from a balance condition between dissipative and inertial
forces acting in the system, and we support this claim through
numerical simulations that allow us for a precise control of the
dissipative forces.

\emph{Experimental setup.}--We study the behavior of a granular system
made of $N=2600$ spheres, with diameter $d=4$ mm, contained in a
cylinder with a conical-shaped floor (diameter 90 mm, minimum height
28.5 mm, maximum height 47.5 mm), see Fig. \ref{fig00}, with
  packing fraction $\sim 49\div52\%$.  The mass of each particle is
  $m=0.267$ g for steel, $m=0.0854$ g for glass, and $m=0.0462$ g for
  delrin.  The container is vertically vibrated by an electrodynamic
  shaker (LDS V450) following the protocol:
\begin{equation}
z(t)=A \sin(2\pi f t) 
\label{zmax}
\end{equation}
where $z$ is the vertical coordinate of the shaker plate. The maximal
acceleration is $\ddot{z}_{max}=A (2\pi f)^2$. The explored frequency
and amplitude ranges are $30\div 700$ Hz and $0.014\div 0.053$ mm,
respectively. Higher values of $f$ cannot be reached in our
setup. Errors on the fixed vibration amplitudes are about $10\%$. A
Plexiglas vane (height 15 mm, width 6 mm, length 35 mm) is suspended
in the medium, and is subject to an external torque. A dc motor
  coupled with the rotator is operated at 3 V, producing a torque
  $\tau\sim 6\times 10^{-3}$Nm, see Supplemental Material
  (SM)~\cite{SM}. Further details on the experimental setup are given
in~\cite{gnoli,scalliet,gnoli2}. On the top of the granular medium we
place a thick aluminum plate (mass $M_{top}=218$g).  The vertical
displacement of the plate $\Delta Z$ can be measured by a laser device
optoNCDT 1400, while the angular position of the vane $\theta(t)$ is
recorded by an encoder.

\begin{figure}[!tb]
\includegraphics[width=0.8\columnwidth,clip=true]{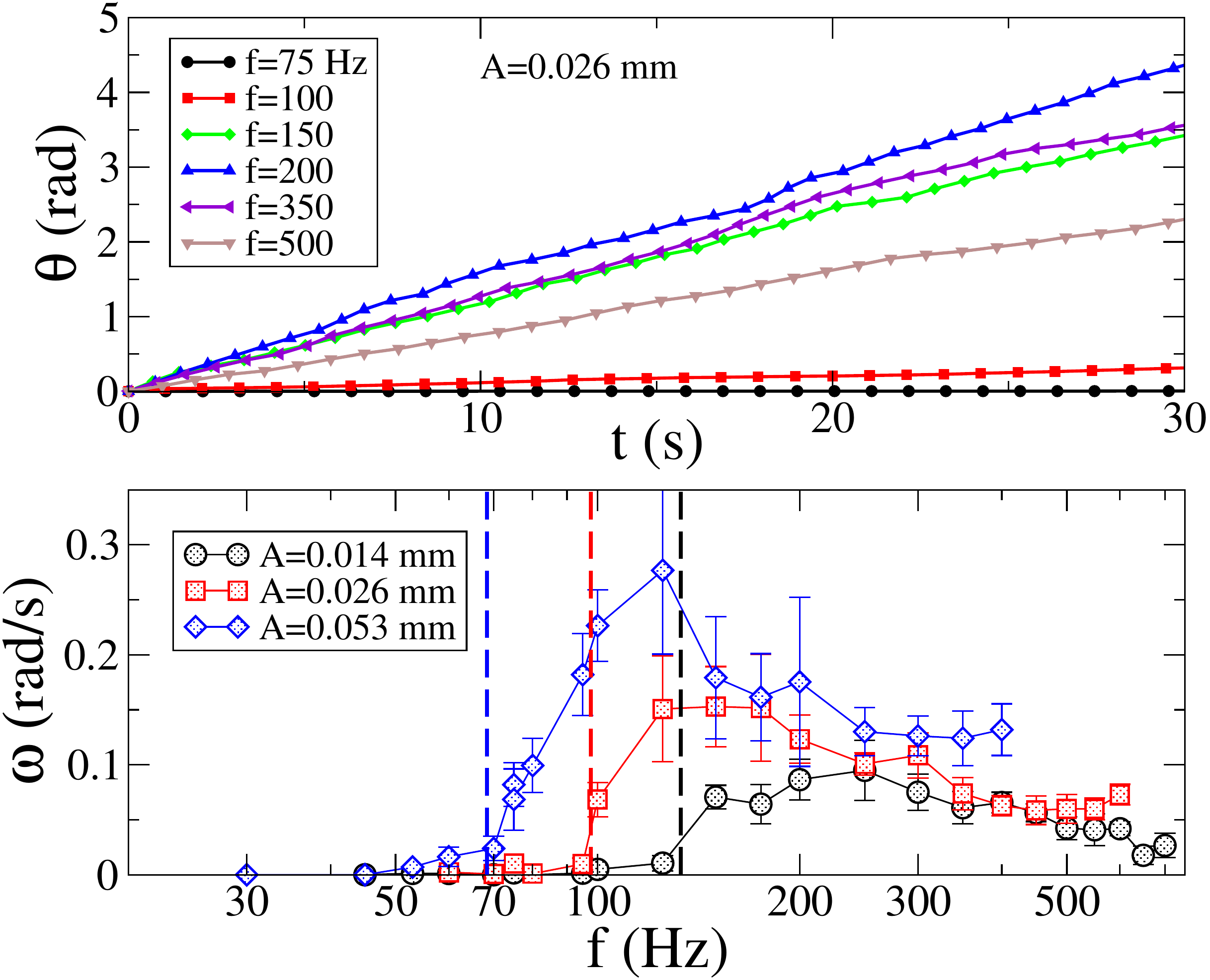}
\caption{Top panel: Angular position $\theta(t)$ of the rotator as a
  function of time, for different values of $f$, at fixed amplitude
  $A=0.026$ mm, for steel spheres. Bottom panel: Average angular
  velocities $\omega$ as a function of $f$ for three shaking
  amplitudes. The vertical dashed lines represent the theoretical
  predictions for the fluidization frequencies, see
  Eq.~(\ref{f1}).}
\label{fig0}
\end{figure}

\emph{Activated fluidization.}-- In the absence of vibrations, the
applied torque is not able to fluidize the system, which is in a
static jammed configuration.  We have considered how fluidization
occurs when we drive the system, investigating the role of $f$, for
some fixed values of $A$.  In Fig.~\ref{fig0} (top panel), we show a
typical time dependence of the rotator angular position for different
experiments with steel spheres, shaken at $A=0.026$ mm.  From the
signal $\theta(t)$ we obtain the average angular velocity
$\omega=\langle d\theta(t)/dt\rangle$, where the average is taken over
trajectories of 30 seconds. Considering the applied torque constant,
the inverse of $\omega$ is proportional to the macroscopic viscosity
of the system.  The values of $\omega$ as a function of $f$, for
different amplitudes $A=0.014,0.026,0.053$ mm are reported in the
bottom panel of Fig.~\ref{fig0}.  The behavior of the system, as
probed by the rotating vane, is characterized by three regimes. First,
at low vibration frequencies, the vane velocity is zero, corresponding
to infinite viscosity. This is due to the low energy fed into the
system: the granular medium remains at rest in its jammed state,
frictionally interacting with the vane. In the second regime, the vane
angular velocity rapidly increases and reaches a maximum value
$\omega_{max}$. This behavior reflects the jammed-unjammed transition,
induced by the mechanical vibrations, and corresponds to a viscosity
reduction in the system. As detailed below, such a fluidized regime
corresponds to the detachment condition from the vibrating substrate,
where the granular medium expands and the top plate reaches its
maximum height. In the third regime, for higher values of $f$,
$\omega$ decreases, signaling an increasing viscosity. Our
experimental setup is similar to the one used in~\cite{danna}, where
the granular fluid was described as a thermalized system, displaying
Brownian motion. In our case, the high density system leads to a more
complex phenomenology~\cite{gnoli}.  Similar studies for granular
suspensions are presented in~\cite{hanotin1,hanotin2}, where a model
predicting their rheology is proposed. However, the dependence on the
vibration frequency is not investigated.

\begin{figure}[!tb]
\includegraphics[width=0.8\columnwidth,clip=true]{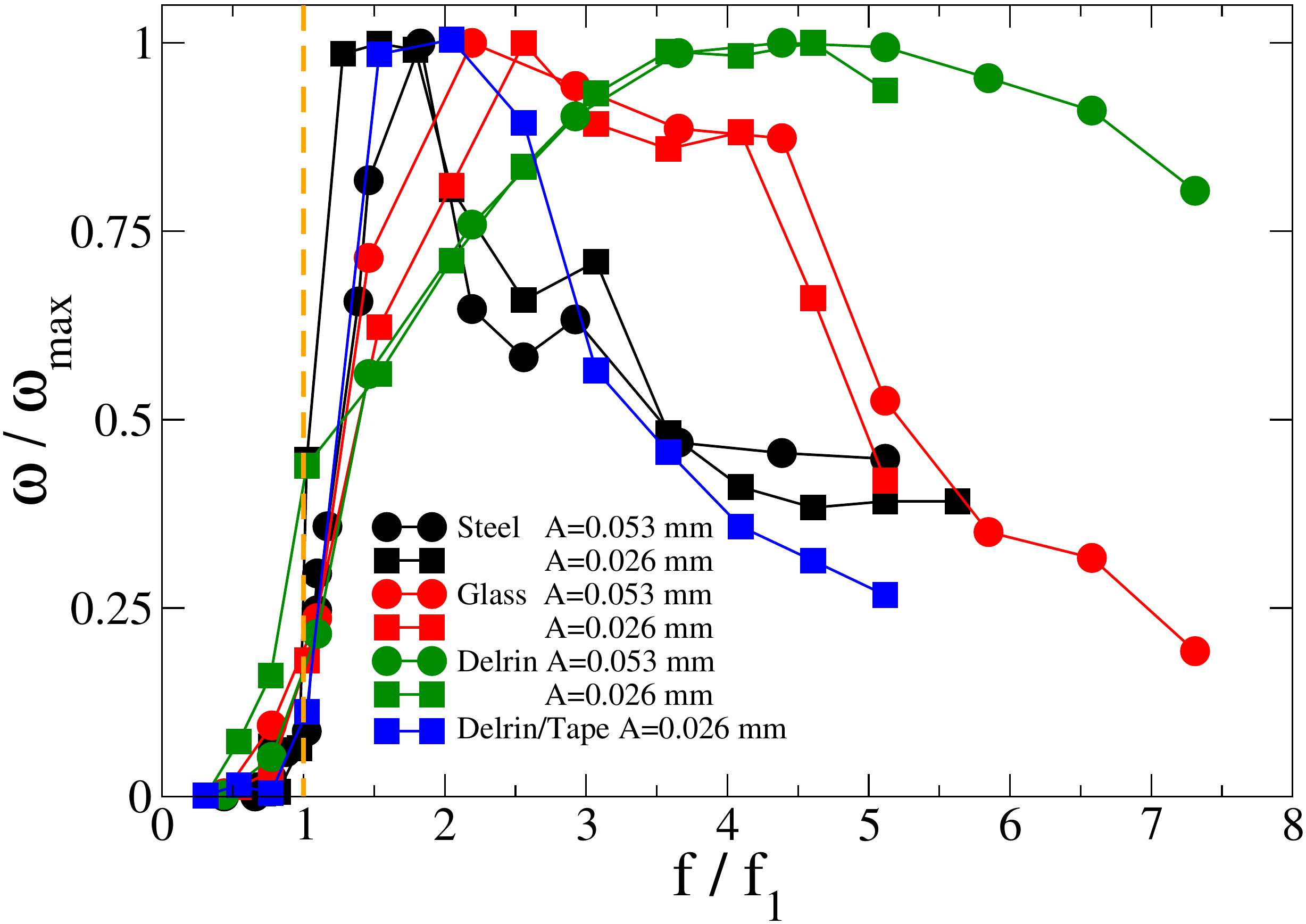}
\caption{Rescaled angular velocity, as a function of $f/f_1$, for
  different materials. The vertical dashed line marks the fluidization
  threshold, according to the theory. Data points are obtained as an
  average over 10 realizations of the same experiments, with standard
  deviation $\sim 15\%$.}
\label{fig2}
\end{figure}

The raise of $\omega$ going from the first to the second regime,
occurs at a well defined value of $f$, that we denote by $f_1$. A
quantitative estimation of this value can be obtained from the
theoretical argument discussed in
Refs.~\cite{capozza,capozza1}. Indeed, the fluidization condition is
realized when the largest force provided by the shaker,
$F=M\ddot{z}_{max}$, where $M$ is the total mass of the system
(granular particles and top plate), equals the weight $Mg$, with $g$
the gravity acceleration.  According to this argument, from
Eq.~(\ref{zmax}), the fluidization frequency $f_1$ is given by the
following relation:
\begin{equation}
2\pi f_1 = \sqrt{g/A}.
\label{f1}
\end{equation}
This expression predicts an explicit dependence of the fluidization
frequency on the vibration amplitude, very well confirmed by our
experimental data, as reported in the bottom panel of Fig.~\ref{fig0}
for steel spheres (see vertical dashed lines).  In order to
demonstrate the generality of the fluidization mechanism, in
Fig.~\ref{fig2} we report data obtained in experiments with different
materials (steel, glass and delrin). We rescale the frequencies
by $f_1$, defined in Eq.~\ref{f1}, and $\omega$ by the maximum value
$\omega_{max}$ for each data set, obtaining a good collapse of the
curves at the onset of the fluidization region. Notice that the
activation frequency does not depend significantly on the material, as
predicted by Eq.~(\ref{f1}). The dependence of $\omega_{max}$ on $A$
is approximately linear, yielding, for steel spheres,
$\omega_{max}=bA$, with $b\simeq 4.66$ s$^{-1}$mm$^{-1}$, as obtained
from data of Fig. \ref{fig0}. Let us note that this
fluidization phenomenon is different from the acoustic fluidization
mechanism~\cite{giaccoprl2015,melosh}, related to acoustic 
waves bouncing back-and-forth within the medium~\cite{note}.

\begin{figure}[!tb]
\includegraphics[scale=0.31,clip=true]{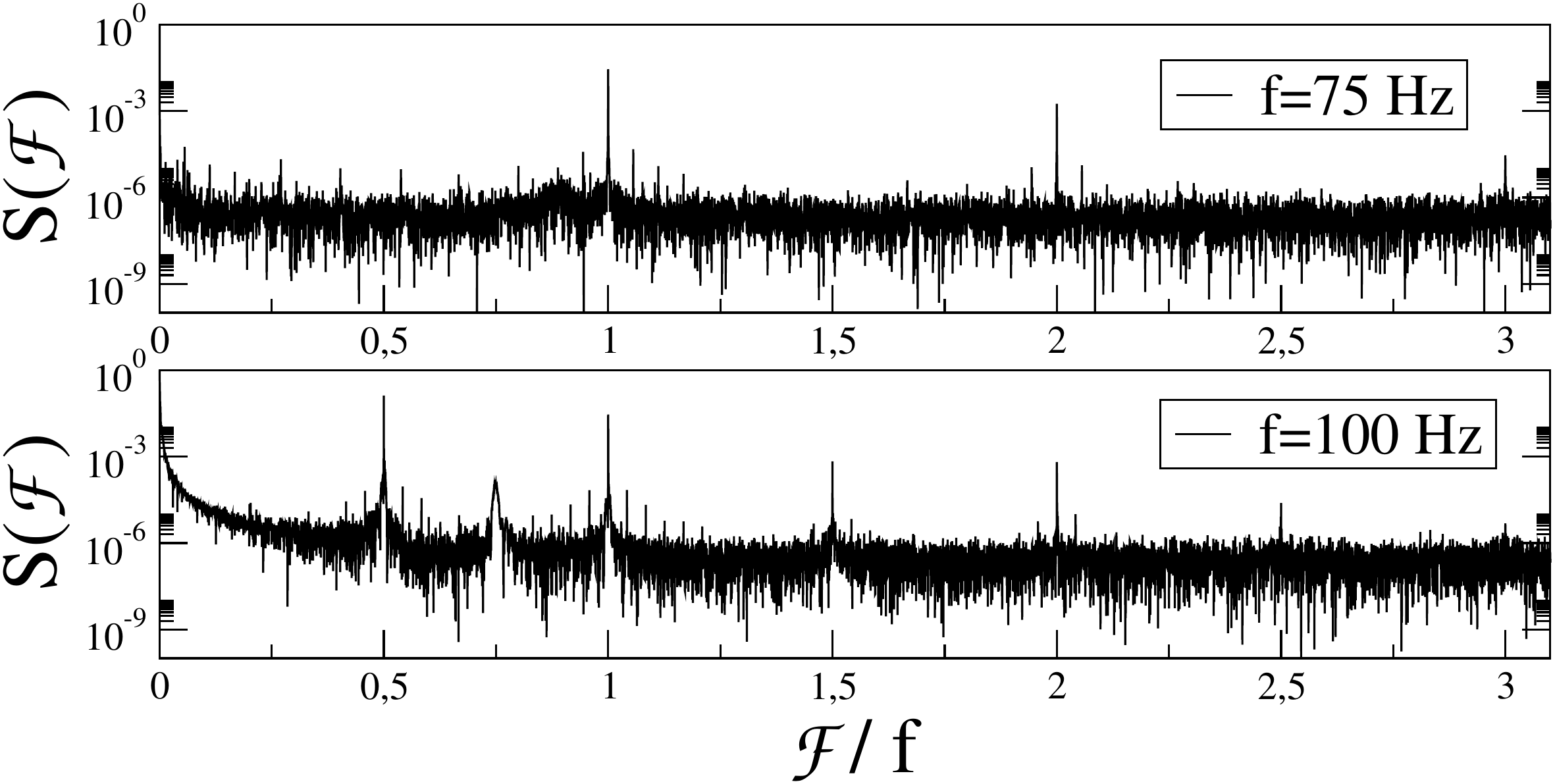}
\caption{Power spectrum $S(\mathcal{F})$ of the signal $\Delta Z(t)$,
  measured in experiments with steel spheres at $A=0.026$ mm, in the
  frictional regime (top panel) and in the fluidized regime (bottom
  panel).}
\label{fig2b}
\end{figure}

The detachment is confirmed by the system dilation for
$f>f_1$. Further insights are provided by the top plate vertical
displacement $\Delta Z(t)$.  In good agreement with what observed in
the numerical simulations reported in Ref.~\cite{capozza}, the power
spectrum $S(\mathcal{F})$ of the signal $\Delta Z(t)$ shows pronounced
peaks at integer multiples of $f$ for $f<f_1$, while in the
fluidization region additional peaks at multiple values of $f/2$ do
appear, see Fig.~\ref{fig2b}. As already observed in \cite{capozza1},
this phenomenology is similar to the problem of period-doubling as a
route to chaos in the bouncing ball and related models
\cite{luck,dewijn}.

\emph{Viscosity recovery.}-- Remarkably, the system exits the state of
minimum viscosity at vibration frequencies $f\gtrsim f_2$, as shown in
Fig.~\ref{fig2}. Viscosity recovery at high frequencies has been
previously observed in the model system of Ref.~\cite{capozza} and in
numerical simulations of a driven spring-block model~\cite{giacco}.
According to the argument of Ref.~\cite{capozza}, the viscosity
recovery is expected to occur when the detachment time from the bottom
plate equals the period of the external oscillation. Since in our
experiments there is no confining pressure on the top plate and the
total normal force is simply $F_N=Mg$, the value of $f_2$ would be
proportional to the fluidization frequency $f_2 =\sqrt{2\pi}f_1$,
without any dependence on the materials.  On the contrary, the
recovery frequency $f_2$ observed in our experiments shows a marked
dependence on the material: $\omega$ drops to about $0.5\omega_{max}$
at frequencies $f_2\sim 3.5 f_1$ and $5f_1$, for steel and glass,
respectively (Fig.~\ref{fig2}).

A similar phenomenology is observed in a simple spring-block model
under vertical vibration~\cite{giacco}. There, the second transition
to a state of larger viscosity originates from a balance between
dissipative and inertial forces. More precisely, according to
Ref.~\cite{giacco}, $f_2$ depends on the dissipation rate. In our
system, this quantity is affected by the elastic and dissipative
forces characterizing the grain-grain and grain-interface
interactions. To clarify the role of the dissipation at the
medium-bottom interface in our system, we performed experiments
where the bottom plate is covered with a thick layer of rubber tape,
reducing the restitution coefficient in the collisions with the
grains. As shown in Fig.~\ref{fig2}, the recovery frequency is
significantly reduced in this case (compare blue squares to green
squares), namely $f_2$ decreases upon increasing the dissipation in
the system.

\emph{Numerical simulations.}-- To confirm the above argument and
obtain a quantitative explanation, we have performed molecular
dynamics simulations of a granular medium of $N=1000$ grains of
unitary mass $m$ and diameter $d$, enclosed between two plates. The
plates are made of closed packed grains whose relative positions are
kept fixed during the dynamics. The system is confined by the
gravitational force and has dimensions $L_x\times L_y=20d\times 5d$,
with periodic boundary conditions along the $x$ and $y$ directions.
We use a standard model for the interparticle interaction, see SM for
details~\cite{SM}.  The vertical distance $L_z$ between the plates is
not fixed, as the system is allowed to expand under vibration, but
typically $L_z \simeq 10d$. Data for larger system size are reported
in the SM~\cite{SM} and show similar behaviors, suggesting that our
results are robust and not a finite size effect.

\begin{figure}[!tb]
\includegraphics[scale=0.31,clip=true]{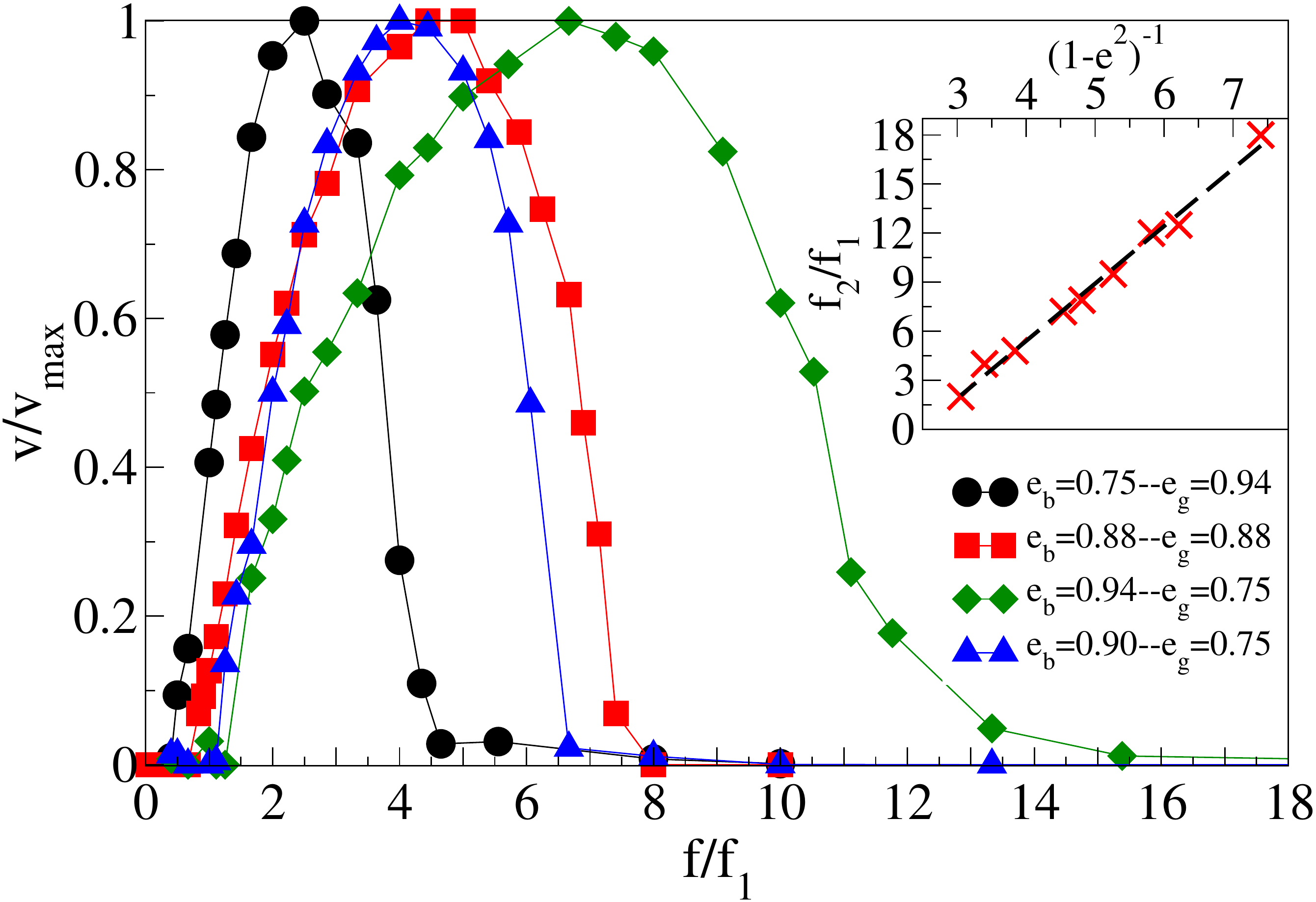}
\caption{{Numerical simulations.} Rescaled velocity of the probe as a
  function of $f/f_1$. Data are obtained for systems with different values
  of $e_g$ and $e_b$, chosen in analogy with the experiments (see
  Fig.~\ref{fig2}). Inset: The rescaled values of the recovery
  frequency $f_2$ as a function of the inverse of the dissipation
  factor $1-e^2$ for systems with $e_g=e_b$.}
\label{fig4}
\end{figure}

The bottom plate moves according to the same protocol used in the
experiments (i.e. $z(t)=A \sin(2\pi f t)$).  To study the viscous
properties of the system, we monitor the motion of a rigid
cross-shaped subset of $5$ grains, touching and glued to each other,
and lying in the plane $z$-$y$. This probe, playing the role of the
vane in the experiment, is subject to a constant force $F$ along the
$x$ direction, while the positions of the $5$ grains are kept fixed
along $z$ and $y$. Time is measured in units of $t_0$ and the
integration step is $5\cdot 10^{-4}t_0$. Other parameters are $F=500
\,md/t_{0}^{2}$ and $g=10\, d/t_{0}^2$.  We employ a contact force
model that captures the major features of granular interactions, known
as linear spring-dashpot model, taking into account also the presence
of static friction, as fully described
in~\cite{giacco,cundall,silbert}.  Our numerical simulations take into
account both normal and tangential frictional forces among grains. We
measure the velocity $v$ of the probe along the $x$ direction
(averaged over trajectories of $3\cdot 10^5t_0$) for different values
of $f$ and $A$, chosen in the range $f\in
[10^{-3}t_{0}^{-1},10^{-1}t_{0}^{-1}]$ and
$A\in[0.05d,0.2d]$. Numerical results show a low frequency
fluidization transition at $f_1$ in good agreement with
Eq.~(\ref{f1}), followed by a viscosity recovery at higher frequencies
(Fig.~\ref{fig4}).  This complex behavior of the velocity is supported
by the non-monotonic behavior of the average particle coordination
number, shown in SM \cite{SM}. A similar behavior of this quantity is
also observed in the range of frequencies of acoustic
fluidization~\cite{olson}.

These results, combined with the analysis of translational and
rotational kinetic energies (see SM~\cite{SM}), indicate that in the
high frequency regime, the system attains high density values, and a
relevant fraction of kinetic energy is rotational, in agreement with
Refs.~\cite{DGJMC17,FGGJMC15}, for systems under shear.  In the
following, we focus on the viscosity recovery transition observed at
higher frequencies whose behavior is expected to depend on the
dissipation mechanisms.

In numerical simulations we can change the viscoelastic properties of
the system by tuning the rigidity of each grain, corresponding to a
change in the restitution coefficient of each grain
$e$~\cite{silbert}. Frictional dissipation can be neglected, see
SM~\cite{SM}. More precisely, to reproduce the experimental setup, we
consider two different restitution coefficients: $e_{g}$ for
grain-grain collisions, and $e_{b}$ for collisions between grain and
bottom plate.  In Fig.~\ref{fig4} we show the value of $v/v_{max}$ as
a function of $f/f_1$ for different values of $e_{g}$ and $e_{b}$.
Results clearly indicate that $f_1$ is not affected by $e_b$ and
$e_g$, whereas the recovery frequency $f_2$ depends on the
dissipation. In particular, we expect that the smaller is the fraction
of energy lost in a collision, the higher is the value of the recovery
frequency $f_2$, since the system needs a larger number of collisions
to dissipate the amount of energy necessary for the viscosity
recovery. More specifically, the rate of energy dissipation can be
estimated as $(1-e^2)f$~\cite{kumaran} (the main assumption here being
that the collision frequency is $\propto f$), giving a condition for
the viscosity recovery of the kind $(1-e^2)f_2 > const$. To verify
this dependence of $f_2$ on $e$, we consider the simplest case
$e=e_g=e_b$, and define $f_2$ as the value of the shaking frequency
for which the velocity of the probe is $v=0.25v_{max}$.  From the
results reported in the inset of Fig.~\ref{fig4} we obtain a behavior
$f_2/f_1 \sim (1-e^2)^{-1}$, in agreement with the above argument.  As
in the experiments, these findings are not consistent with the
scenario of~\cite{capozza}, where the frequency $f_2$ is related to
the rise time associated with the internal vibrations of the grains.
Our numerical results also indicate that the main dissipation in the
system occurs at the medium-bottom interface, as illustrated in
Fig~\ref{fig4}, where black and green symbols correspond to sets with
$e_b$ and $e_g$ values interchanged. This scenario is confirmed by the
direct measurement of the energy dissipated in the bulk and at the
bottom interface, see SM~\cite{SM}. Note that in the simulations we
did not consider a dependence of the restitution coefficient on the
impact velocity. Since the collision velocity changes with the
vibration frequency, the restitution coefficient could increase up to
30\%, depending on the material
parameters~\cite{kuwabara,brilliantov,ramirez,mcnamara}. This could
explain some differences between the experimental and numerical
curves.

\emph{Conclusion.}-- We have studied experimentally the viscous
properties of dense granular materials under vertical vibration. Using
a vane subject to an external torque as probe, we have observed
different regimes in the system, from very large viscosity at low
vibration frequencies, to fluidized states (corresponding to viscosity
reduction) at intermediate $f$, with a viscosity recovery at higher
values of $f$. The first transition to the fluidized state is well
characterized by the detachment condition, and is independent of the
material properties. The second transition, leading to viscosity
recovery, turns out to be related to dissipation mechanisms in the
medium and between the medium and the bottom plate, and therefore
shows a strong dependence on materials.  Our study suggests the
possibility to control the viscous properties of confined granular
media by tuning the shaking frequency in the system, with important
practical application in several fields, from tribology to geophysics
and material industry.

\emph{Acknowledgments.}-- This paper is dedicated to the memory of our dear friend and colleague
Ferdinando Giacco, who passed away too prematurely.
He made the time spent together special,
in and out of the office, with his critical thinking, irony, and kindness.





\onecolumngrid

\newpage

\begin{center}

\Large{\textbf{Supplemental Material}}

\end{center}

\section{Rheology experiment}

\begin{figure}[h!]
\includegraphics[width=6cm]{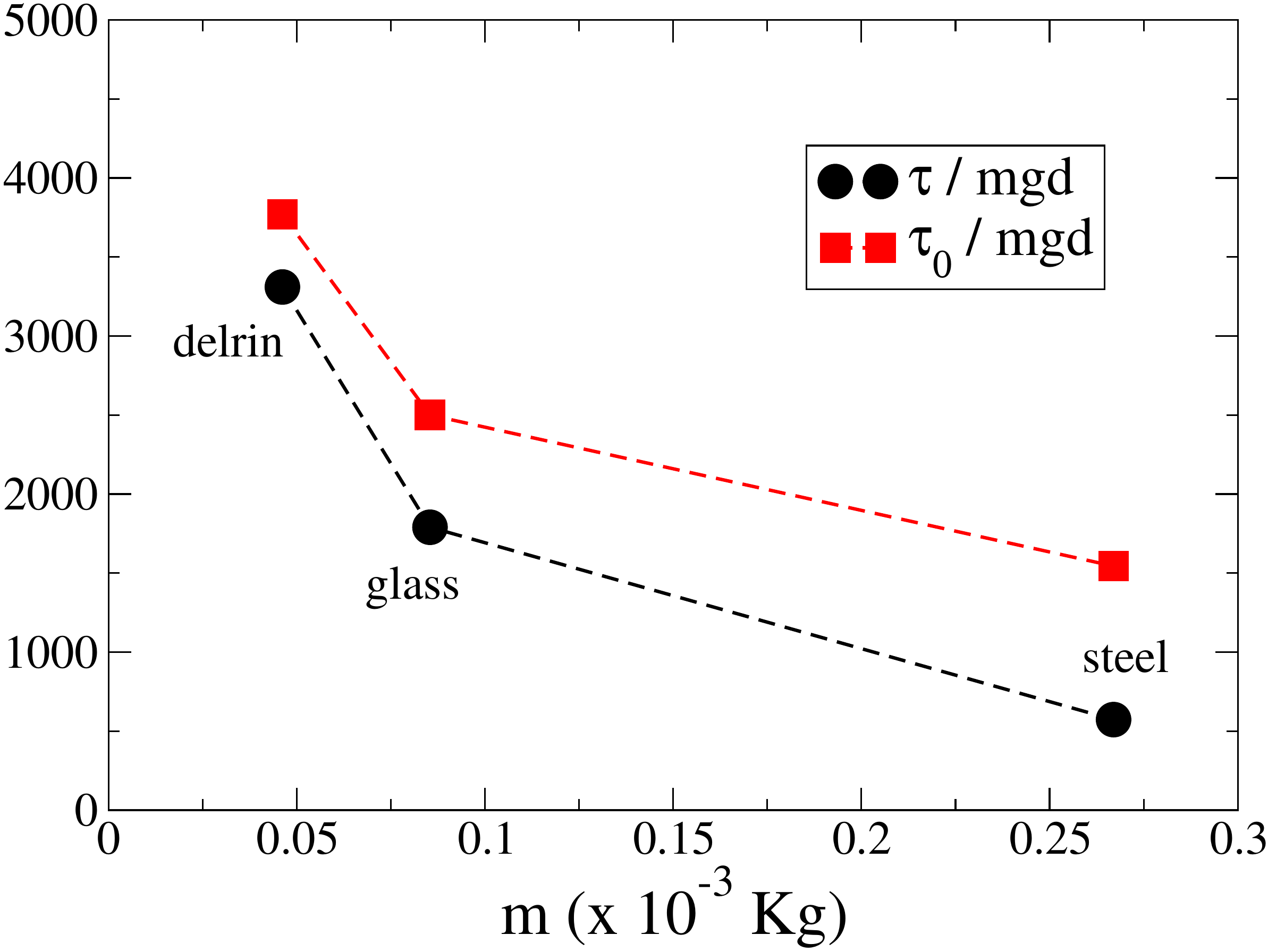}
\caption{Dimensionless ratio $\tau/mgd$ (black circles) and lower
  bound $\tau_0/mgd$ (red squares) for $\mu=0.5$ for different materials.}
 \label{s0}
\end{figure}

In our experimental setup we probe the viscosity features of the
granular system measuring the average angular velocity of a vane
suspended in it and driven by a dc motor.  The motor is operated at 3
V, producing a torque $\tau\sim 6\times 10^{-3}$ Nm. In Fig.~\ref{s0}
we show the dimensionless ratio $\tau/mgd$, where $m$ is the mass of
one particle, $g$ the gravity acceleration and $d$ the particle
diameter, for different materials. To give a physical meaning to the
values obtained for the ratio $\tau/mgd$, we estimate a lower bound
for the torque required to move the vane through the granular medium
(in the absence of external vibration). For simplicity, we consider a
disc with radius $R$ and height $l$ equal to those of the vane, and estimate the
tangential frictional forces acting on the disc surface due to the medium.
Then, the minimum torque necessary to rotate the disc is given by
$\tau_0\sim R S \mu P$, where $R=17.5$ mm,
$S$ is the surface of the disc ($2\pi Rl=1650$ mm$^2$, where $l=15$ mm, we neglect the upper and lower faces of the disc), $\mu$ a friction coefficient and $P$ the pressure acting on the disc.  The
pressure $P$ can be estimated as $P\sim \phi \rho g h+Mg/\pi a^2$, where
$\phi$ is the packing fraction ($\sim 50\%$) of the granular medium,
$\rho$ the material density ($\rho=0.008$ g/mm$^3$ for steel, $\rho=0.0025$ g/mm$^3$ for glass and $\rho=0.0014$ g/mm$^3$ for delrin), $g$ the gravity acceleration, $h$ the
depth of the disc in the medium ($\sim 20$ mm), $M$ the mass of the top plate ($218$
g), and $a$ the radius of the container ($45$ mm). Then, we obtain that $\tau/mgd$
is smaller than the lower bound
$\tau_0/mgd$, for realistic values of $\mu\geq 0.3$.
This is
consistent with our observation that in the absence of vibration the
system remains in a jammed state, because the vane alone cannot
fluidize the medium.
In the figure we compare the ratio
$\tau/mgd$ to $\tau_0/mgd$ for $\mu=0.5$ for all materials. 

\section{Details of the numerical model}

We have performed molecular dynamics simulations of spherical
frictional grains. We use a standard model for the interparticle
interaction, detailed as model L2 in L. E. Silbert, D. Ertas,
G. S. Grest, T. C. Halsey, D. Levine, and S. J. Plimpton Phys. Rev. E,
64, 051302, (2001).  Two particles of diameter $d$ only interact when
in contact, i.e. when the relative distance between their centers is $r < d$ and the
overlap $\delta = d-r > 0$.  In this case, the force has a normal and
a tangential component.  The normal component is
\[ \vec F_n = (k_n \delta  - \gamma_n m \dot \delta) \vec n\]
where $k_n$ is the stiffness of the particle, $m$ the effective mass,
$\gamma_n$ a normal damping coefficient.  The restitution coefficient
$e$ of this model does not depend on the impact velocity, and is a
function of $k_n$ and $\gamma_n$:
\begin{equation}
e =
\exp[-\pi/(2\sqrt{2k_n/(m\gamma_n^2)-1/4})].
\end{equation}
In our simulations, we have fixed $\gamma_n=50/t_0$ ($t_0=1$) and tuned the restitution coefficient
acting on $k_n$, as experimentally we have considered particles with
different stiffness.  Specifically, we have introduced a restitution
coefficient $e_g$ for particles in the bulk, and a restitution
coefficient $e_b$ for the interaction between a particle in the bulk
and one forming the rough bottom of the container.

The tangential interaction force is given by
\[ \vec F_t = k_t \vec \delta_t,\]
where $k_t$ is a tangential stiffness and $\vec \delta_t$ is the tangential
shear displacement, which is defined as the integral of the relative
velocity at the point of contact. However, to implement Coulomb's
condition, at every integration timestep the magnitude of $\delta_t$
is fixed to $\mu |F_n|/k_t$ ($\mu=0.5$) if $|F_t| > \mu |F_n|$. 
In our simulations we use $k_t=(2/7) k_n$
as usual assumed in the literature. In the fluidized regime,
contacts continously form and break and therefore
the average value of $\delta_t$ is much smaller than the
average overlap $\delta$. For this reason dissipation due
to tangential friction is negligible.

\section{Dissipated energy}

\begin{figure}[h!]
\includegraphics[width=6cm]{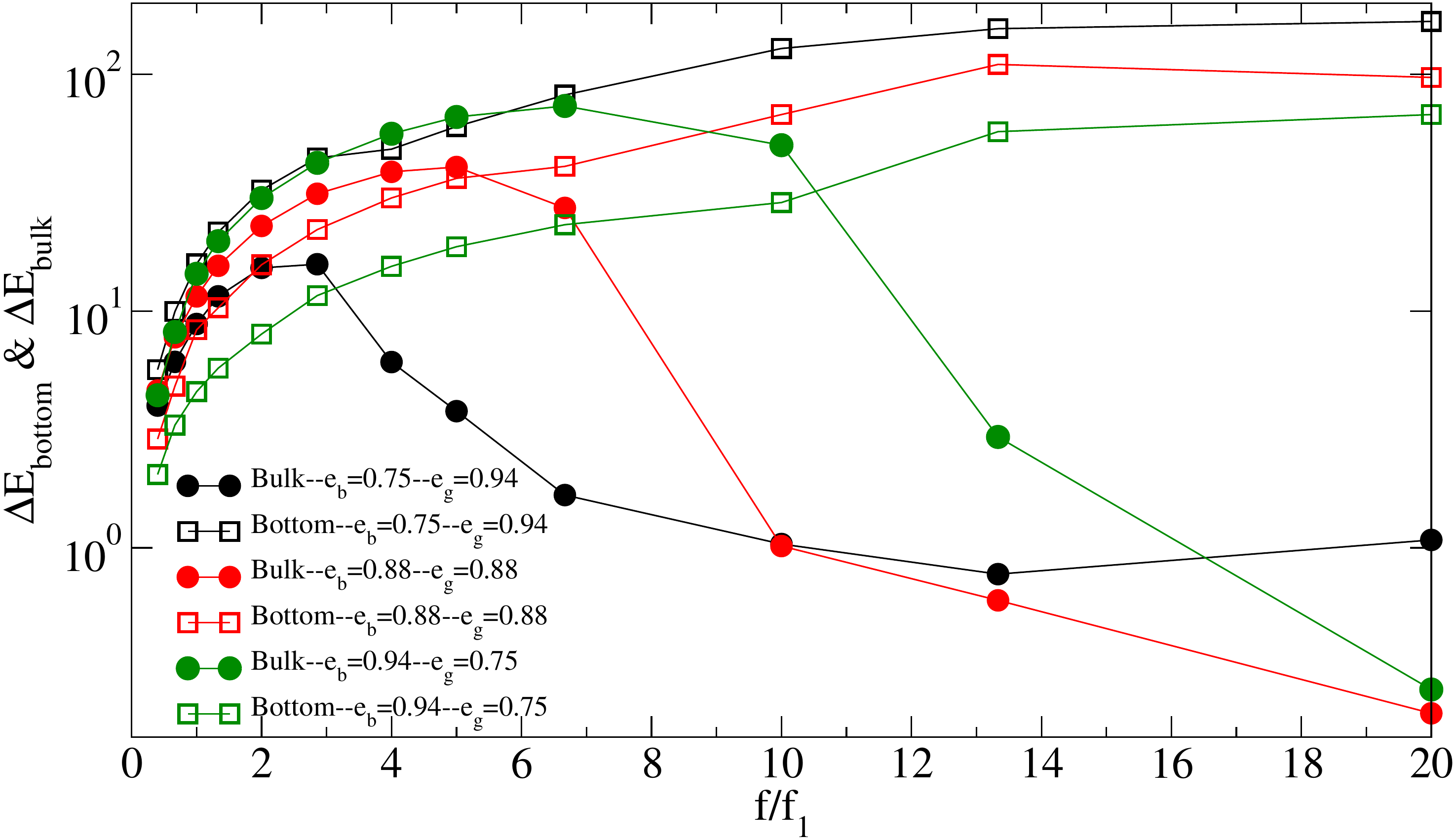}
\caption{In numerical simulations we report the energy dissipated per unit time in
  the interaction between the bottom plate with the granular material,
  $\Delta{E}_{bottom}$ (open squares), and in the
  bulk of the granular material, $\Delta{E}_{bulk}$ (filled
  circles). Different colors correspond to different elastic
  properties of the grains.}
 \label{s1}
\end{figure}

In Fig.~\ref{s1} we show the energy dissipated per unit time, measured in
numerical simulations, at the interface between the bottom plate and
the granular material, $\Delta E_{bottom}$, and in the
bulk of the medium, $\Delta{E}_{bulk}$. Results show that for $f\gtrsim f_2$, interaction
  with the boundary becomes the dominant dissipation mechanism. This
  effect becomes more relevant the softer are the bottom-plate grains.

\section{Coordination number}

\begin{figure}[h!]
\includegraphics[width=5cm]{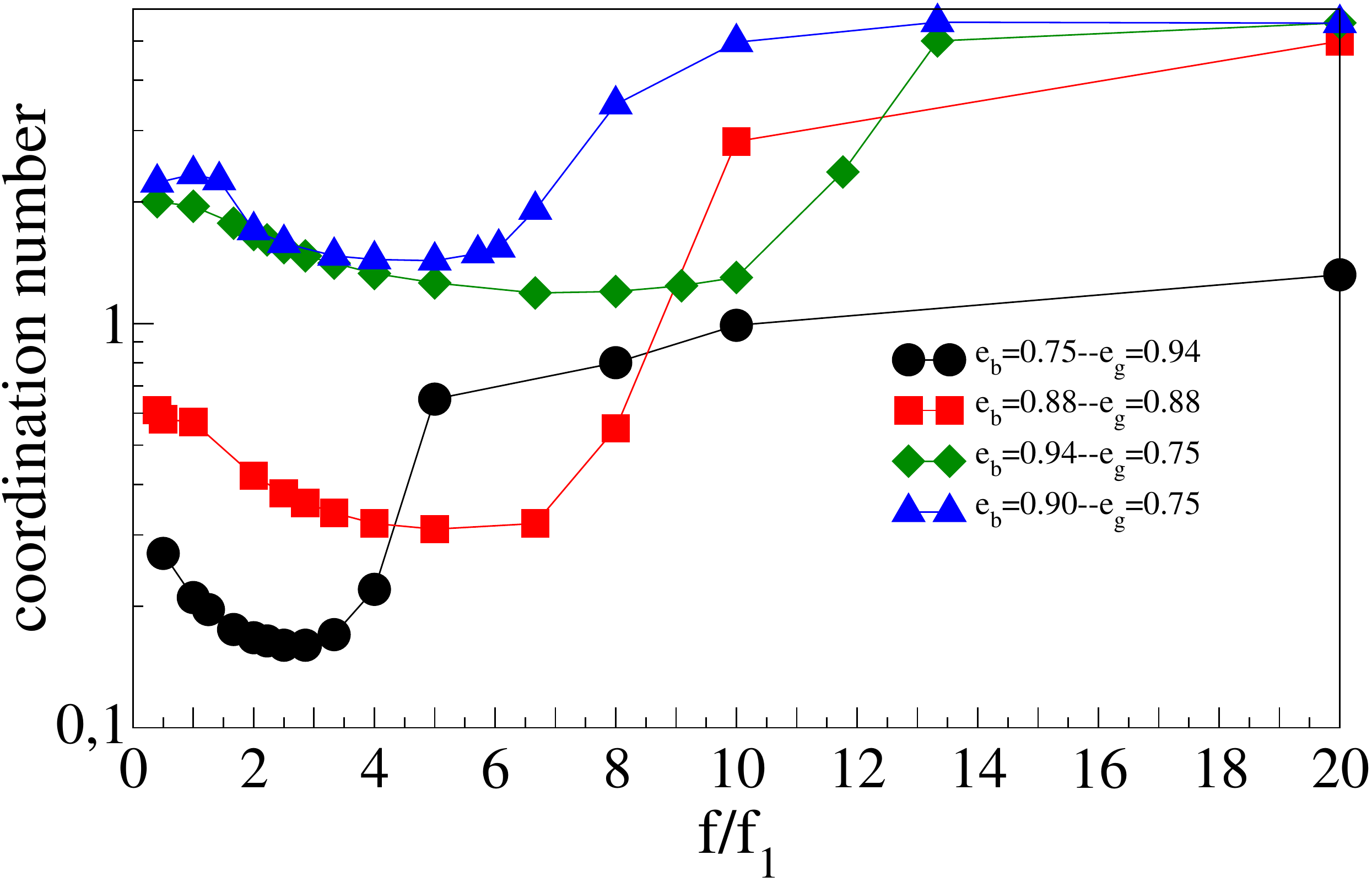}
\caption{Coordination number measured in numerical
  simulations. Different colors and symbols are used for the different
  elastic properties of grains in the bulk as well as in the
  bottom plate, with the same color and symbol code of Fig.5 of the manuscript.}
 \label{s2}
\end{figure}

In Fig.~\ref{s2} we report the coordination number defined as the
average number of particles at distance smaller than one diameter from
the center of a given particle. At low frequencies, softer grains 
present a higher coordination number. In all cases, for
increasing $f$, the coordination number presents a non-monotonic
behavior with a minimum at the frequency $f^{max}$ which corresponds
to the frequency with maximum velocity in Fig.5 of the manuscript.  It
is interesting to observe that the two curves (green diamonds and blue
triangles) having the same value of $e_g=0.75$, but different $e_b$,
are very similar up to $f \simeq 4f_1$. This indicates that at small
frequencies the dissipation with the bottom plate does not play a
crucial role. At larger frequencies, conversely, curves separate
drastically with grains in the system with the softer bottom plate
presenting a larger coordination number. Results suggest that
the system approaches a dense state at high frequencies where no translation is possible.

\newpage
\section{Translational and rotational kinetic energy}

\begin{figure}[h!]
\includegraphics[width=6cm]{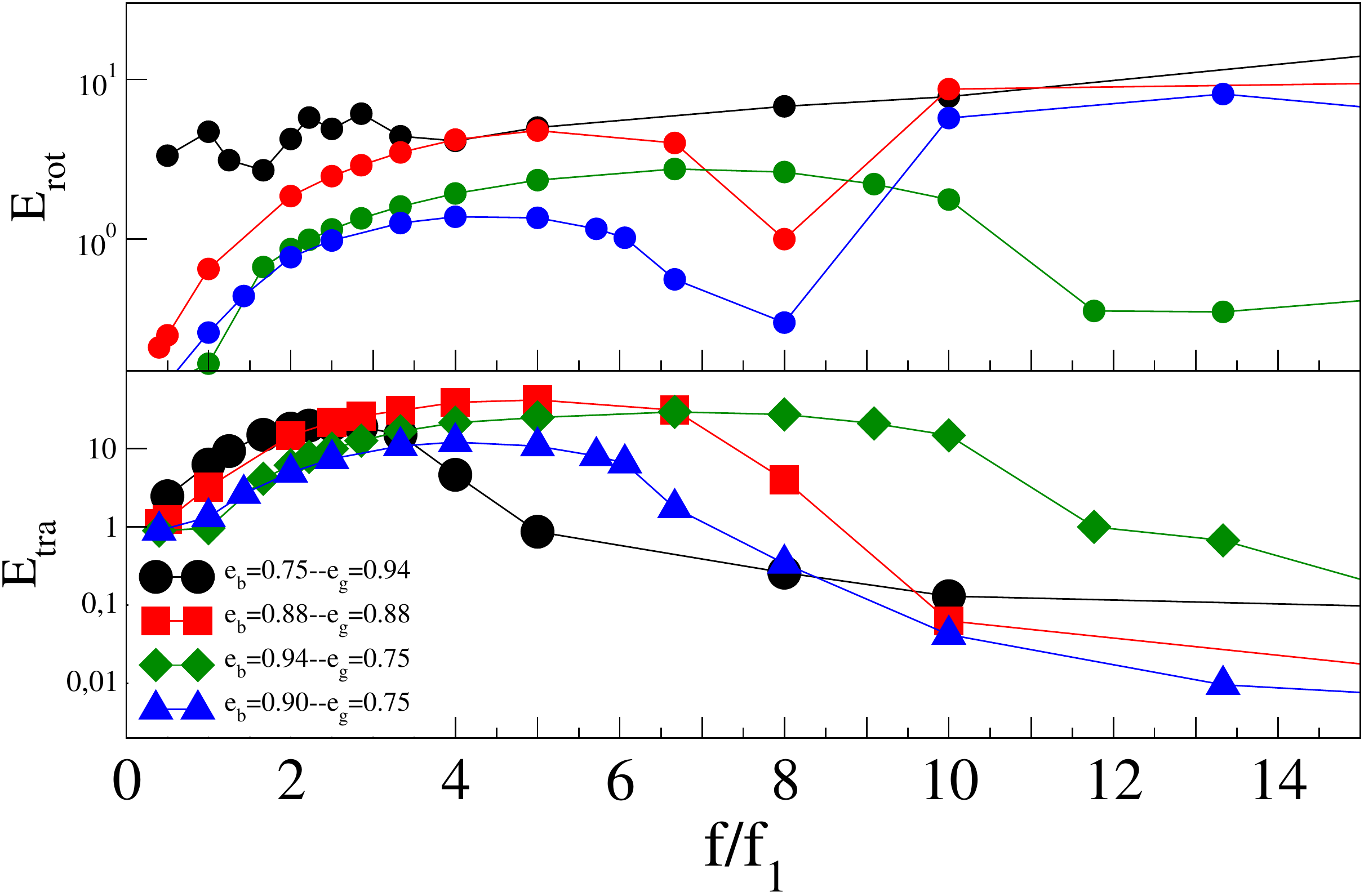}
\caption{In numerical simulations, we measure the rotational
  energy per particle $E_{rot}$ (upper panel) and the translational kinetic
  energy per particle $E_{tra}$ (lower panel), as a function of $f/f_1$.  The same
  color and symbol code of Fig.5 of the manuscript is used.}
\label{s3}
\end{figure}

In Fig.~\ref{s3} we report the rotational kinetic energy per particle,
$E_{rot}$, and the translational kinetic energy per particle, $E_{tra}$. The
rotational energy presents a non trivial behavior, as function of $f$:
An initial increase for $f \lesssim f^{max}$ followed by a decay for $f^{max}
\lesssim f \lesssim f_2$ and a final increase for $f \gtrsim f_2$.  Also
the translational energy presents a non-monotonic behavior, with a
maximum value of the ratio $E_{rot}/E_{tra}$ at $f = f^{max}$. At
larger frequencies $f>f^{max}$, $E_{tra}$ decreases quickly to zero
indicating that for $f>f_2$ translational motion is suppressed. The
increase of $E_{rot}$ for $f>f_2$, conversely, indicates that, even if
the system is approaching a denser state, the energy injected by the
vibrating plate amplifies the grain rotation.

\section{Finite size effects}

\begin{figure}[h!]
\includegraphics[width=6cm]{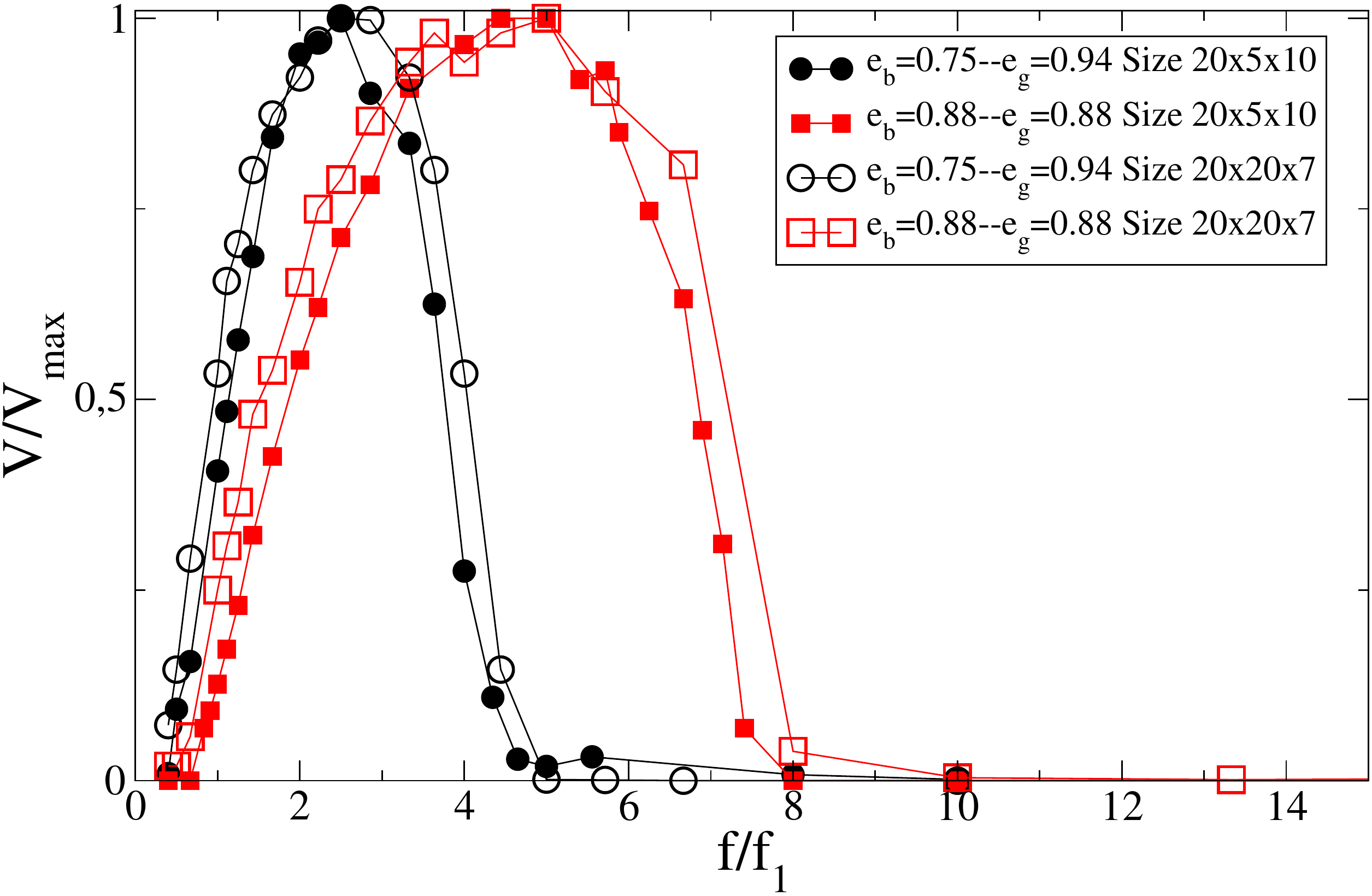}
\caption{Numerical simulations for systems with different sizes.}
\label{s4}
\end{figure}

We study how the system size affects the dependence of viscous friction on the
frequency of the vibrating plate. We simulate a system larger (see legend of Fig.~\ref{s4}) than the
one presented in the manuscript, which better corresponds to the
experimental set up.  Results indicate only a weak dependence on the
system size.

\end{document}